\documentclass[apjl]{emulateapj}
\usepackage[latin2]{inputenc}
\usepackage{amsmath}
\usepackage{graphicx}
\usepackage{url}
\usepackage{multirow}

\begin{document}
\title{A Cold Milky Way Stellar Stream in the Direction of Triangulum}
\author{Ana Bonaca\altaffilmark{1}, Marla Geha\altaffilmark{1}, Nitya
Kallivayalil\altaffilmark{1}} 
\altaffiltext{1}{Department of Astronomy, Yale University, New Haven, CT 06511; {ana.bonaca@yale.edu}}

\begin{abstract}
  We present evidence for a new Milky Way stellar tidal stream in the
  direction of the Andromeda and Triangulum (M31 and M33) galaxies.
  Using a matched-filter technique, we search the Sloan Digital Sky
  Survey Data Release 8 by creating stellar density maps which probe
  the Milky Way halo at distances between $8$ and $40$\,kpc.  A visual
  search of these maps recovers all of the major known stellar
  streams, as well as a new stream in the direction of M31/M33 which
  we name the Triangulum stream.  The stream spans $0.2^\circ$ by
  $12^\circ$ on the sky, or 75\,pc by 5.5\,kpc in physical units with
  a best fitting distance of $26\pm4$\,kpc.  The width of the stream
  is consistent with being the tidal remnant of a globular cluster.
  A color magnitude diagram of the stream region shows an overdensity
  which, if identified as a main sequence turn-off, corresponds to an
  old ($\sim12$\,Gyr) and metal-poor ([Fe/H]$~\sim-1.0$ dex) stellar
  population. Future kinematic studies of this and similar cold
  streams will provide tight constraints on the shape of the Galactic
  gravitational potential.
\end{abstract}
\maketitle

\section{Introduction}

Stellar streams are found throughout the galactic halos of the Milky
Way, Andromeda (hereafter M31) and other nearby massive galaxies
\citep[e.g.,][]{belokurov2006, ibata2001, md2008}. These streams
are presumed to originate from disrupting dwarf galaxies or globular
clusters, providing evidence that galaxies are formed, at
least in part, hierarchically.

The recent discovery of numerous Milky Way stellar streams is due to
wide-field photometric surveys such as the Sloan Digital Sky Survey (SDSS).
Many of these streams are associated with known objects such as the
tidal tail system of the Sagittarius dwarf spheroidal galaxy
(\citealt{majewski2003}, \citealt{belokurov2006}) or with globular
clusters (e.g., Palomar~5; \citealt{odenkirchen2003}, Palomar~14;
\citealt{sollima2011}). Some streams, however, are identified as
isolated features in which the progenitor object is not yet identified
or has been tidally disrupted beyond recognition (e.g., Monoceros
Stream, \citealt{newberg2002}; Orphan Stream,
\citealt{belokurov2007}).

Stellar streams can extend over large physical distances and are
sensitive to the strength and shape of the Galactic gravitational
potential \citep[e.g.,][]{helmi2004,johnston2005,lux2012}, as well as
to its 'lumpiness' \citep{carlberg2009, yoon2011}. The GD-1 tidal
stream \citep{gd} is sufficiently nearby (10\,kpc) that proper motions
are available for its member stars. \citet{koposov2010} analyzed the
full, six--dimensional phase space for GD-1, placing a strong
constraint on the flattening of the Milky Way potential between
10-15\,kpc which is dominated by the Galactic disk. An analysis of
more distant streams will provide similar constraints on the Galactic
halo potential. Each individual stream provides independent
constraints on the Galactic potential, thus a joint analysis of many
streams provides a promising method to constrain the full Milky Way
potential.

Here we present evidence for a new Milky Way stellar stream, which we
name the Triangulum stream, found in projection against M31
and M33. In \S\,\ref{sec:dataanalysis} we discuss the SDSS data
analysis and matched-filter technique.  In \S\,\ref{sec:results}, we
determine the stream properties, and discuss its relationship to other
Galactic substructures in \S\,\ref{ssec:other}.

\section{Data Analysis}
\label{sec:dataanalysis}

We search for stellar streams in the Sloan Digital Sky Survey Data
Release 8 \citep[hereafter SDSS DR8]{dr8}. We select stars brighter
than $r=22$ over the 14,555 deg$^2$ SDSS coverage, of which 5200
deg$^2$ is newly released imaging in the Southern Galactic hemisphere.
At these magnitudes, the majority of stars belong to the Milky
Way disk.  Stars in substructures
typically belong to different stellar populations as compared to the
Milky Way disk and can be distinguished by their chemical composition
and age.

\citet{rockosi2002} first introduced the matched-filter technique in
the context of Milky Way substructure searches.  Matched filtering
works by preferentially weighting stars, in color-magnitude space,
with a higher probability of being part of a stream population.
Filters used in this work are based on the globular cluster M13, whose
old age and low metallicity ($13.5$\,Gyr and [Fe/H$]=-1.6$;
\citealt{vandenberg2000}) are similar to other known cold stellar
streams.  M13 is observed by SDSS, thus no filter transformations are
required.

\begin{figure*}
\begin{center}
\includegraphics[scale=0.95]{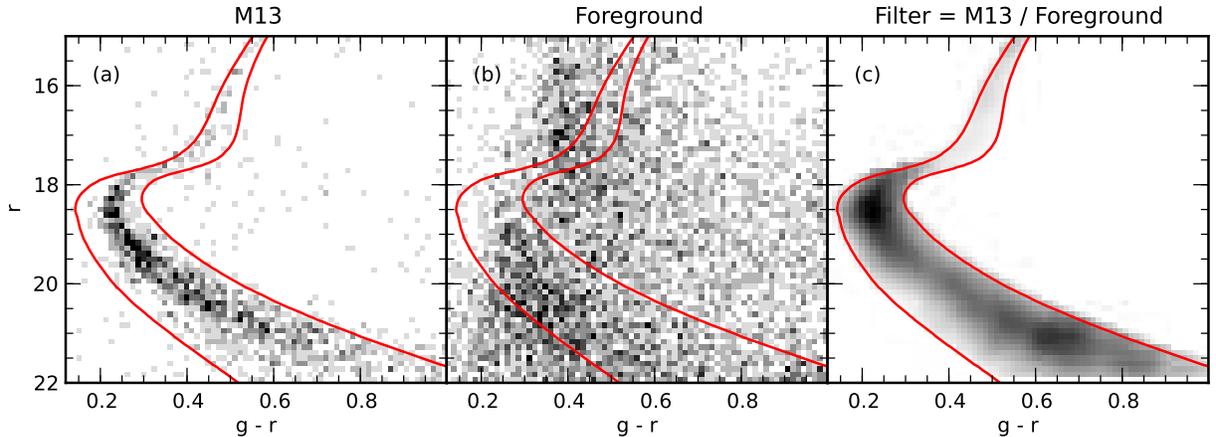}
\caption{Steps in creating an M13-based filter for selection of old, metal-poor
stars:  the left panel shows a Hess diagram of SDSS stars detected in the inner
region of globular cluster M13, while that of a surrounding area, representative
of Milky Way foreground, is given in the middle panel. A ratio of the two
produces the desired filter, which preferentially selects stars that are in the
region of the color-magnitude diagram with the greatest contrast between M13 and
Milky Way stellar populations. In order to reduce clumpiness, the filter is
smoothed (right panel). The resulting filter is most heavily weighted near the
main sequence turn-off. Red lines, derived from an isochrone matching the M13
population, trace the width of the filter on all three panels.}
\label{filter}
\end{center}
\end{figure*}

Our filter is constructed by dividing a Hess diagram of the target
population (Figure~\ref{filter}a) by a Hess diagram of the foreground
(Figure~\ref{filter}b).  The filter target stars are selected from the
DR8 photometric catalog within a radius of $15'$ from M13's
center. The foreground population is selected locally, between $45'$
and $75'$ from M13.    The final filter is smoothed and confined to a
region following the target population's isochrone within a generous
tolerance (Figure~\ref{filter}c), which accommodates isochrones of other
commonly used globular clusters (e.g., M92). 

Our approach ensures that regions in the color-magnitude diagram (CMD)
with high contrast between the target and foreground population are
given the most weight in filtering.  The filter is most heavily
weighted near the main sequence turn-off (Figure~\ref{filter}c), where
numerous M13 stars are present, but few Galactic stars are as blue in
color.  The main sequence is less weighted because of the large
population of Milky Way stars present redward of the turn-off. The
subgiant and giant branches are weighted the least because of the
small number of M13 stars which occupy those areas of the CMD.

\begin{figure*}
\centering
\includegraphics[scale=0.90]{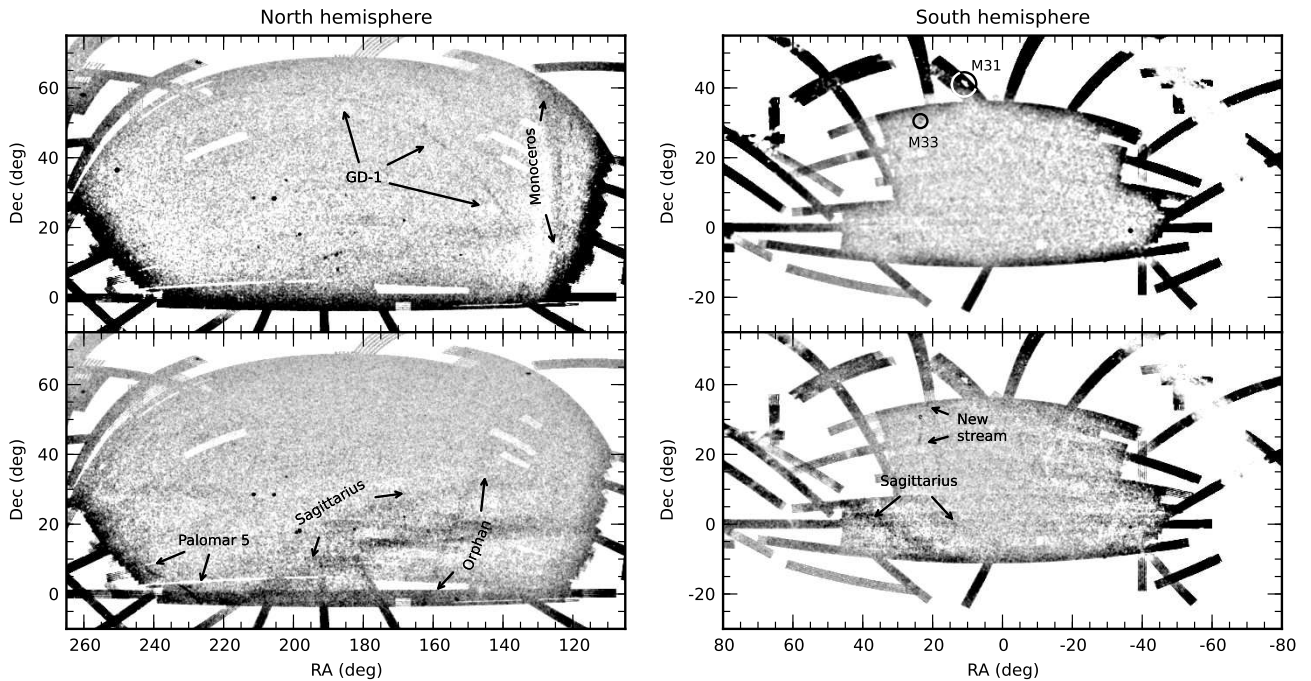}
\caption{Stellar density maps obtained via matched filtering at distances of 8\,kpc (top) and 26\,kpc (bottom).  The left columns show the SDSS footprint in the
northern Galactic hemisphere, while maps on the right are in the south. The
stretch is linear, with darker areas corresponding to higher stellar density
regions. The most prominent features are tidal arms of the Sagittarius dwarf
galaxy, bright Milky Way globular clusters, and brightest neighboring galaxies.
Regions not imaged by the SDSS are shown in white. The maps are available
online.}
\label{map}
\end{figure*}

A total of four matched filters (in $r$ vs.~$g-r$, $g$ vs.~$g-r$, $g$
vs.~$g-i$ and $g$ vs.~$g-z$) were applied to the whole SDSS survey
area for a distance range between 8\,kpc and 40\,kpc. The inner
distance limit was chosen to avoid the Milky Way disk, while the outer
limit is the distance at which the main sequence turnoff falls below
the SDSS 95\% completeness limit \citep{juric2008}.  Each filter was
applied to the full SDSS photometric sample and the resulting density
maps were coadded.  The left column of Figure~\ref{map} shows the final
density maps for the North Galactic hemisphere, while right panels
show the South Galactic hemisphere\footnote{Our SDSS DR8 stellar
  density maps are available at
http://www.astro.yale.edu/abonaca/research/halo.html}. The SDSS
footprint in the North has been extensively searched for halo
substructure and is known as the ``field of streams''
\citep{belokurov2006}.  We recover both compact and diffuse known
objects in this field.  Globular clusters and dwarf galaxies are
visible as small, concentrated overdensities.  At a distance of 8 kpc
(Figure~\ref{map}, top) we detect the cold GD-1 stream, as well as the
much wider signatures of the Monoceros stream and leading Sagittarius
tidal arm. At a distance of 26 kpc (Figure~\ref{map}, bottom) we detect the
tidal tails
of the Palomar 5 globular cluster and the Orphan stream. A visual
search of these maps does not reveal any new stellar streams beyond
those which have been published.  Our method recovers streams down to
the limit of the faint Lethe and Styx streams \citep{grillmair2009}.

Stellar density maps of the South Galactic hemisphere are shown in 
the right column of Figure~\ref{map}, at distances of 8 kpc (top) and 26
kpc (bottom).  The Sagittarius trailing tidal arm is visible at both
distances.  At larger latitudes we also detect a fainter stream parallel to the
stronger Sagittarius tail, as discussed in \citet{koposov2012}.  A visual search
of the Southern maps reveals a new
stream in the region of M31/M33 which we describe in detail below.
No other substructure features were detected.

\section{Results}\label{sec:results}

We demonstrate in Figure~\ref{map} that our search method recovers all
 known stellar streams thus far published.  In addition, we find
evidence for a new narrow stream in the direction of the M31/M33
region which we name the Triangulum stream. In \S\,\ref{ssec:other}, we discuss
this new stream in the context of other known substructure in this region. 

\subsection{Physical Properties of the Triangulum Stream}

\begin{figure*}
\begin{center}

\includegraphics[height=85mm]{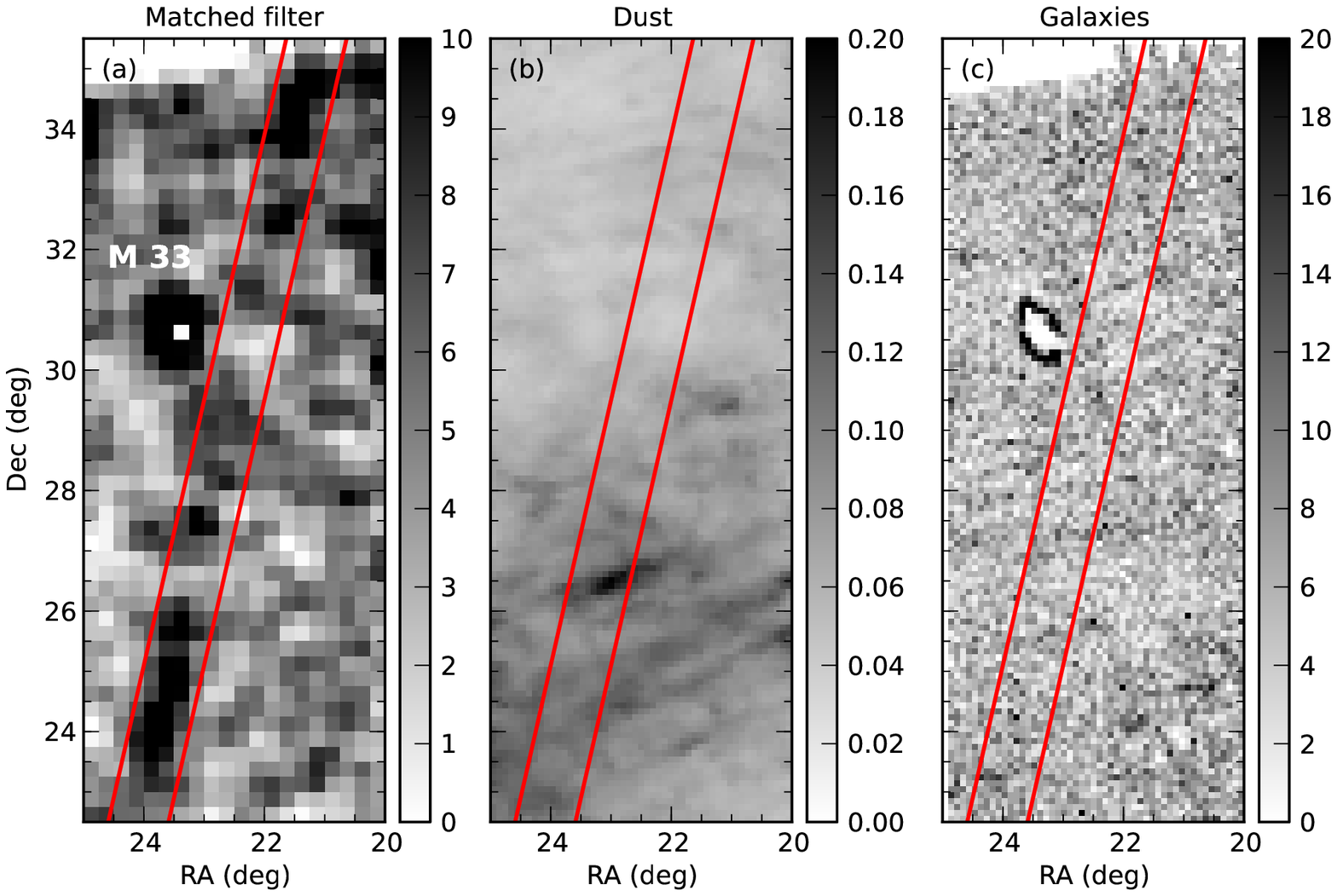}
\quad
\includegraphics[trim = 0mm 0mm 0mm 0mm, clip, height=85mm]{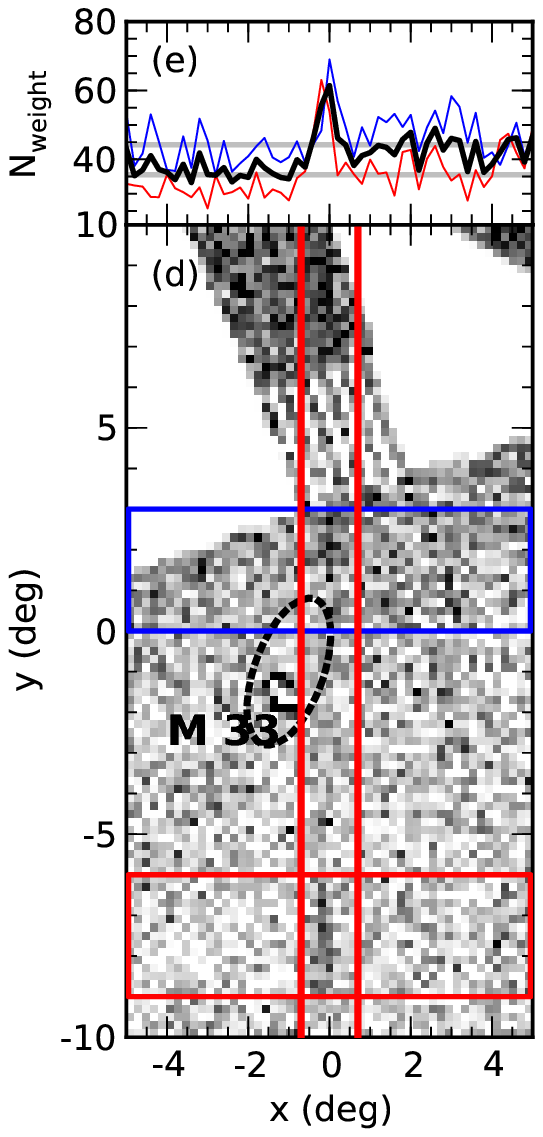}

\caption{
(a) Matched filter map of the Triangulum stream region in equatorial
coordinates. (b) Dust map of the same region. Color scale represents reddening
$E(B-V)$. (c) Blue galaxy number counts in the stream region. (d) Matched filter
map of the wider stream region in rotated coordinates. Stream regions are boxed
in blue and red, the dashed ellipse shows the extent of RGB stars around the M33
disk. Red lines on panels a-d bracket stream region. (e) 1D profiles of the
stream in the top and bottom boxes from the panel (d) (blue and red,
respectively) and their average (black).}


\label{profile}
\end{center}
\end{figure*}

\begin{table}
\caption{Triangulum stream properties}
\label{properties}
\begin{center}
\footnotesize{
\begin{tabular}{l c}
\hline\hline
Property & Value\\
\hline
RA (deg)\footnotemark[1] &  [21.346, 23.975]\\
Dec (deg)\footnotemark[1] &  [34.977, 23.201]\\
$l$ (deg) & [130.756, 136.023]\\
$b$ (deg) & [$-$27.378, $-$38.538]\\
Width & $0.2^\circ$ (75\,pc) \\
Length & $12^\circ$ (5.5\,kpc) \\
Distance & $26\pm4$ kpc \\
$[$Fe/H$]$ & $-$1.0 dex\\
Age & 12 Gyr \\
\hline
\end{tabular}
\footnotetext[1]{Stream boundary points, see Equation~\ref{coords} for a
relation between RA and Dec.}
}
\end{center}
\end{table}

The Triangulum stream is long and very narrow (Figure~\ref{profile}a). To
ensure that the feature is indeed a stellar overdensity, we show the dust
extinction \citep[Figure~\ref{profile}b;][]{schlegel1998} and number counts of
blue SDSS galaxies (Figure~\ref{profile}c) in the stream region. Neither dust
nor galaxies show an overdensity in the stream region.

The Triangulum stream is nearly linear on the sky, whose coordinates satisfy the
equation:
\begin{equation}
\rm Dec (^\circ) = - 4.4\times \rm RA (^\circ) + 128.5
\label{coords}
\end{equation}
in the range RA$\simeq21^\circ-24^\circ$ (Table~\ref{properties}). We define a
new coordinate system such that the stream runs along a $y$-axis.  This
system is an equatorial system rotated $11^\circ$ counterclockwise
around RA $=22^\circ$ and Dec $=32^\circ$. A stellar density map of
the stream region in this new coordinate system is shown in
Figure~\ref{profile}d.  To assess the statistical
significance of this overdensity, we create a one--dimensional stellar
density profile of the stream (regions $-9^\circ<y<-6^\circ$ and
$0^\circ<y<3^\circ$, marked with red and blue lines on
Figure~\ref{profile}d, respectively) extending $|x|\leq 5^\circ$ on
each side of the stream.    For $x<-1^\circ$ the stream profile
is affected by the incomplete SDSS coverage. We correct for
completeness by dividing the profile value with the fraction of
populated pixels in each vertical bin. This procedure enables easier
analysis of the stream profile, but it also introduces additional
noise to the profile at $x<-1^\circ$. The resulting profile is shown
in Figure~\ref{profile}e: blue and red lines are profiles of
top and bottom stream regions, respectively, while the black line is
their average.

The stream signal was strongest for the matched filter at 26 kpc. We
show this quantitatively by fitting a Gaussian function with a
constant term to the stream profile at a range of distances between
20\,kpc and 30\,kpc. A Levenberg--Marquardt minimization algorithm was
used to solve for the mean, dispersion, and normalization of the
distribution. Both the dispersion and peak of the profile are
maximized at a distance of $\sim26$\,kpc, while the mean does not
depend on the distance. The height of the stream peak is $3\,\sigma$
above the background noise (estimated in the region $0.5^\circ<|x|<5^\circ$) at
distances $23-30$\,kpc. We adopt a distance to the stream of $26\pm4$\,kpc.
The measured angular extent is $\Delta~x=10'$ and $\Delta~y=12^\circ$,
which correspond to a width of $75$\,pc and a length of $5.5$\,kpc in
physical units.

We also detected a $\sim2\,\sigma$ overdensity of blue horizontal branch
(BHB) stars in the lower part of the stream region
($-9^\circ<y<-6^\circ$). In this box, the $0.5^\circ$ wide strip
contains on average $7.5\pm0.7$ BHBs in the stream fields
($|x|\leq0.5^\circ$) and $4.6\pm2.1$ BHBs in the off fields
($0.5^\circ<|x|<5^\circ$). 

In general, the shape of a stellar stream traces the past orbit of its
progenitor.  The stream width correlates with progenitor mass, while
the length of a stream is related to the time since the stars became
unbound \citep{Johnston2004}.  The average width of the Triangulum
stream, $\sigma=10\arcmin$ (75\,pc), is comparable to other globular cluster streams in the literature.
The GD-1 stream, whose progenitor is undetected but presumed to be a
globular cluster, is $\sigma = 12\arcmin$ \citep[35\, pc;][]{koposov2010},
while the globular cluster Pal 5 is $\sigma = 18\arcmin$ (120\,pc;
\citealt{odenkirchen2003}).  In contrast, dwarf galaxies produce much
broader streams: the Orphan stream is $\sim 2^{\circ}$ wide
\citep[730\,pc;][]{grillmair2006b, belokurov2006}, while the Sagittarius
stream is $20^{\circ}$ or more
\citep[7\,kpc;][\citealt{majewski2003}]{belokurov2006}.  The width of the
Triangulum stream suggests that its progenitor is a globular cluster, consistent
with the age and metallicity determined below.


Our estimated length for the Triangulum stream, $12^{\circ}$
(5.5\,kpc), is likely an underestimate given the incompleteness of the
SDSS data.  This is shorter, but comparable to the Palomar 5 stream
($22^{\circ}$, 9\,kpc).  Deeper imaging along the Triangulum stream
may reveal a longer extent than detected in the SDSS data.

\subsection{Age and Metallicity of the Triangulum Stream}

We estimate the age and metallicity of the Triangulum stream using the
best-fitting distance of $26\pm4$\,kpc. Due to the small number of stars
associated with the stream, the CMD in this region is dominated by
Milky Way foreground stars.  We therefore follow \citet{koposov2010} in
constructing a statistically subtracted CMD. First, we create a 1D
profile of all stars satisfying the color-magnitude cut $20<r<22$,
$0.2<g-r<0.5$ in the stream region ($|x|<5^\circ$,
$-9^\circ<y<-6^\circ$ and $0^\circ<y<3^\circ$). This produces a well
populated profile, with predominantly stream stars. Next, we fit a Gaussian to
this profile to determine the central position $x_0$ and width $\sigma_0$. To
construct a statistically-subtracted Hess diagram, we
create a 1D profile for stars in every pixel in the Hess diagram, fit
the profile with a Gaussian of central position $x_0$ and width
$\sigma_0$, and assign each pixel the fitted Gaussian
normalization. Thus, higher-value pixels correspond to regions with higher
density profile in the stream region and, consequently, higher probability of
being occupied by a stream rather than a foreground star.

\begin{figure}
\begin{center}
\includegraphics[scale=0.28]{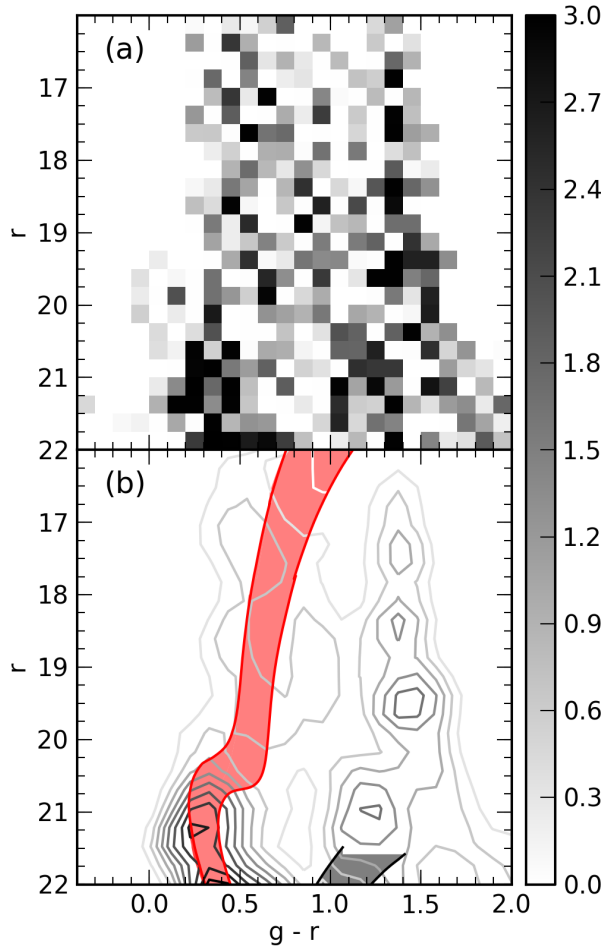}
\caption{Top: Hess diagram of the stream region with statistically subtracted
background. Bottom: Contours of the smoothed Hess diagram with overplotted 12
Gyr isochrones spanning a metallicity range of [Fe/H$]=-0.5$ to $-1.5$ at the
stream distance (red) and [Fe/H$]=-1.1$ to $-2.1$ at the M33 distance (gray).}
\label{cmd}
\end{center}
\end{figure}

The statistically subtracted $r$ vs.~$g-r$ Hess diagram is noisy, but clearly
shows an overdensity at $r>20.5$ and $g-r<0.5$ (Figure~\ref{cmd}a). At a
distance of 26 kpc, this overdensity corresponds to main sequence stars,
and a hint of a main sequence turnoff is indeed seen. The contour of a
smoothed Hess diagram is overplotted with a range of Dartmouth
isochrones \citep[][red isochrones on Figure~\ref{cmd}b]{dotter2008}. The
isochrones are 12 Gyr old and at a distance of
26 kpc. The left-most isochrone is the most metal-poor, [Fe/H$]=-1.5$,
while the right-most is the most metal rich, [Fe/H$]=-0.5$. These values
bracket the likely stream metallicity range, with the most probable value being
the median [Fe/H$]=-1.0$. If a range of 13 Gyr isochrones is used, the
inferred metallicity is lower [Fe/H$]=-1.25$. A range of younger and
more metal rich isochrones fails to match the shape of the overdensity in the
Hess diagram. 

Given the proximity of the stream to the M33 galaxy, we have also overplotted
a range of old and metal-poor isochrones at the distance of M33 \citep[809
kpc;][]{mcconnachie2005} to test whether the stream could be associated with the
M33 system (gray isochrones on Figure~\ref{cmd}b). The
tip of the red giant branch of that isochrone only reaches $r=21.5$ at
$g-r\simeq1$, much redder than the stream feature. We conclude that
the stream is not associated with M33, but is most likely the tidal
remnant of an old, metal-poor globular cluster in the Milky Way
halo.

\subsection{Relationship to Other Substructures}\label{ssec:other}

The Triangulum stream is in a region of sky rich with substructure.
While the stream is projected against the M31/M33 system, it has
no spatial overlap with any of the M31 satellites
\citep{richardson2011} or diffuse structures discovered around M33
(marked with a dashed ellipse in Figure~\ref{profile}d).  The M33
population is more metal-poor ($[\rm Fe/H]=-1.6$;
\citealt{mcconnachie2010}) than the Triangulum stream.  Thus, we
believe that Triangulum stream is a Galactic feature, unassociated
with the substructure in the M31/M33 system.

The largest Milky Way feature in this region is the diffuse
Triangulum-Andromeda (TriAnd) overdensity covering $\sim1000$ deg$^2$
between $100^\circ<l<150^\circ$, $-20^\circ>b>-40^\circ$. It was
discovered by \citet{rocha-pinto2004} as a smooth overdensity of
M-giants in the 2MASS survey between
$15-30$\,kpc. \citet{majewski2004} refined this distance estimate
using main sequence turnoff stars to $16-25$\, kpc.  The mean
spectroscopic metallicity of TriAnd stars is [Fe/H$]=-0.64\pm0.08$,
with a CMD inferred age of 8\,Gyr \citep{chou2011}.  The metallicity
of TriAnd is higher than that inferred for the Triangulum stream and
we estimate a population which is 4\,Gyr older.  \citet{martin2007}
have detected a signature of another overdensity behind the TriAnd, at
heliocentric distance $\sim28$\,kpc.  Both features are far more
extended than Triangulum, however, the Triangulum stream could be
related to these overdensities as a localized feature at the outer
edge of a much larger diffuse merger remnant.  Spectroscopic and/or
astrometric follow-up is required to explore this possibility.

Recently, it has been argued that the `bifurcation' in the Sagittarius
stream \citep{belokurov2006} is in fact a distinct stream with its own
progenitor and kinematics \citep{koposov2012}. The so-called Cetus
stream, first discovered by \cite{newberg2009}, is fainter, thinner
and more metal-poor than the Sagittarius stream.  Its derived orbit is
in the direction of the Triangulum stream, however, it has lower
metallicity ([Fe/H$]=-2.1$) and appears to be at larger heliocentric
distance \citep[34\,kpc;][]{newberg2009}. Similarly, we do not think
that
the Triangulum Stream is related to the Monoceros Ring/Anticenter
Stellar Structure/Monoceros Stream, which is the other prominent
diffuse structure in the South. Even though the latter structure
(whose origin remains controversial) has a large metallicity spread
from [Fe/H$]=-1.6$ to [Fe/H$]=-0.4$ that could be consistent with the
stream discovered here, it seems to be at much closer heliocentric
distance at $\sim 11$ kpc \citep{yanny2003,crane2003,chou2010}.


\section{Conclusions}

We have searched the SSDS DR8 photometric database
for Milky Way stellar streams using a matched filtering technique. We recover
all known streams in the current literature.  Based on a visual search of our
stellar density
maps, we present evidence for a new stream discovered in the southern
Galactic hemisphere in the direction of M33 which we name the
Triangulum stream. We did not find any other stream-like structures in the
dataset. 

The Triangulum stream spans $0.2^\circ$ by $12^\circ$ (75\,pc by
5.5\,kpc) on the sky with a best fitting distance of $26\pm4$\,kpc.
We estimate the age and metallicity of the stream via a statistically
subtracted color-magnitude diagram of the region, which suggests a
12\,Gyr, metal-poor ([Fe/H$]=-1.0$) stellar population. The
narrow width of the Triangulum stream is comparable to the GD-1 and
Palomar~5 streams, from which we conclude that the Triangulum stream
is also a disrupted globular cluster.  While we do not observe the
progenitor object, the stream extends beyond the SDSS photometric
coverage. We note that the PAndAS team \citep{mcconnachie2009}
has independently detected the Triangulum stream at the same sky position but at
a higher significance due to deeper photometric data (M.~Fardal, priv.
communication).

Searches for cold stellar streams in the halo are motivated by the
desire to map the Galactic gravitational potential.  Further
characterizing the size and shape of the Triangulum stream via deeper
and more extended imaging will improve the usefulness of this stream
as a Galactic halo probe.  Radial velocities and proper motions of
Triangulum stream stars will provide tighter constraints on the nature
of its stellar populations and orbital properties, providing new
constraints on the Galactic gravitational potential.

\section*{Acknowledgments}

The authors wish to thank Jeremy Bradford, Nhung Ho,
Mario Juri\' c and Beth Willman for providing useful comments on the
manuscript. We also thank Nikhil Padmanabhan for help in the early stages of
this project. This work was supported by NSF grant AST-0908752 and the Alfred
P.~Sloan Foundation. 

Funding for SDSS-III has been provided by the Alfred P. Sloan Foundation, the
Participating Institutions, the National Science Foundation, and the U.S.
Department of Energy Office of Science. The SDSS-III web site is
\url{http://www.sdss3.org/}.

SDSS-III is managed by the Astrophysical Research Consortium for the
Participating Institutions of the SDSS-III Collaboration including the
University of Arizona, the Brazilian Participation Group, Brookhaven National
Laboratory, University of Cambridge, Carnegie Mellon University, University of
Florida, the French Participation Group, the German Participation Group, Harvard
University, the Instituto de Astrofisica de Canarias, the Michigan State/Notre
Dame/JINA Participation Group, Johns Hopkins University, Lawrence Berkeley
National Laboratory, Max Planck Institute for Astrophysics, Max Planck Institute
for Extraterrestrial Physics, New Mexico State University, New York University,
Ohio State University, Pennsylvania State University, University of Portsmouth,
Princeton University, the Spanish Participation Group, University of Tokyo,
University of Utah, Vanderbilt University, University of Virginia, University of
Washington, and Yale University. 

\bibliographystyle{apj}

\end{document}